\documentclass[preprint,pra,aps,showkeys,showpacs,longbibliography]{revtex4-1}
\usepackage{ucs}
\usepackage[english]{babel}
\usepackage{amsmath}
\usepackage{amssymb,amsfonts}
\usepackage{color}

\usepackage{graphicx}
\usepackage{dcolumn}
\usepackage{bm}
\usepackage[utf8]{inputenc}
\usepackage[T1]{fontenc}
\usepackage{tikz}
\usetikzlibrary{shapes.misc}
\usepackage{mathrsfs}
\usepackage[T1]{fontenc}

\newcommand{\ew}[1]{\big\langle #1 \big\rangle}
\newcommand{\ndg}{{\phantom{\dagger}}}
\newcommand{\dg}{\dagger}
\newcommand{\ket}[1]{| #1 \rangle}
\newcommand{\bra}[1]{\langle #1 |}

\begin{document}

\preprint{APS/123-QED}

\title{Quantum cascade driving: Dissipatively mediated coherences}

\author{Shahabedin C. Azizabadi}
\email{shahab@ut.ee}
\affiliation{Technische Universit\"at Berlin, Institut f\"ur Theoretische 
Physik, Nichtlineare Optik und Quantenelektronik, Hardenbergstra{\ss}e 36, 10623 
Berlin, Germany}
\author{Nicolas L. Naumann}%
\affiliation{Technische Universit\"at Berlin, Institut f\"ur Theoretische 
Physik, Nichtlineare Optik und Quantenelektronik, Hardenbergstra{\ss}e 36, 10623 
Berlin, Germany}
\author{Manuel Katzer}%
\affiliation{Technische Universit\"at Berlin, Institut f\"ur Theoretische 
Physik, Nichtlineare Optik und Quantenelektronik, Hardenbergstra{\ss}e 36, 10623 
Berlin, Germany}
\author{Andreas Knorr}
\affiliation{Technische Universit\"at Berlin, Institut f\"ur Theoretische 
Physik, Nichtlineare Optik und Quantenelektronik, Hardenbergstra{\ss}e 36, 10623 
Berlin, Germany}
\author{Alexander Carmele}
\affiliation{Technische Universit\"at Berlin, Institut f\"ur Theoretische 
Physik, Nichtlineare Optik und Quantenelektronik, Hardenbergstra{\ss}e 36, 10623 
Berlin, Germany}
\date{\today}

\begin{abstract}
Quantum cascaded systems offer the possibility to 
manipulate a target system with the quantum state of
a source system.
Here, we study in detail the differences between  
a direct quantum cascade and coherent/incoherent driving for 
the case of two coupled cavity-QED systems.
We discuss qualitative differences between these
excitations scenarios, which are particular strong for higher-order 
photon-photon correlations: $g^{(n)}(0)$ with $ n>2 $.
Quantum cascaded systems show a behavior differing from the
idealized cases of individual coherent/incoherent driving
and allow to produce qualitatively different
quantum statistics.
Furthermore, the quantum cascaded driving exhibits 
an interesting mixture of quantum coherent 
and incoherent excitation dynamics. 
We develop a measure, where the two regimes intermix 
and quantify these differences via experimentally 
accessible higher-order photon correlations.
\end{abstract}

\pacs{Valid PACS appear here}
\maketitle

\section{Introduction}
Quantum light sources are realized for 
many different material platforms in semiconductor,
atom and molecular systems \cite{RIS_5,PhysRevLett.109.113601,RIS_4,RIS_7,LPOR:LPOR201000039}  and offer
an exciting testbed for nonlinear quantum dynamics \cite{PhysRevB.84.125324},
including quantum ghost imaging, 
two-photon-spectroscopy \cite{1367-2630-15-3-033036,PhysRevLett.118.030501,
RevModPhys.88.045008}  and quantum light spectroscopy 
\cite{RIS_0,PhysRevA.73.013813,KIRA2006155}.
Prototypical single photon emitters based on semiconductor
nanostructures  are produced \cite{RIS_6,Gaisler2009,0034-4885-75-12-126503}  and used in quantum cryptography protocols \cite{RIS_8,zoller2005quantum,PhysRevLett.84.4729}  and quantum sensing \cite{Walmsley525}.
Recently, practical realization of intense and tunable 
thermal sources have become accessible \cite{PhysRevLett.53.663,PhysRevB.93.241306,PhysRevA.84.053806,jahnke2016giant} 
 and are applied experimentally
for photon-statistics excitation spectroscopy
 \cite{PhysRevLett.115.027401,PhysRevB.93.241306} and to read-out quantum beating
of hyperfine levels via a modulation with pulse 
separation \cite{PhysRevLett.53.663}.
Polarization-entangled photon sources, another class 
of quantum light sources, are electrically driven and triggered
on demand \cite{RIS_1}. 
For highly-
efficient and indistinguishable twin photon sources \cite{RIS_2} 
this is possible as well in the context of N-photon bundle emitters
\cite{RIS_4} and on demand time-ordered 
photon pairs \cite{muller2014demand} .
The rich variety of quantum light is accompanied by exciting
proposals.
Single photon excitation purifies non-classical states and 
suppresses fluctuations \cite{PhysRevLett.115.196402}  and allows for 
Hilbert-state addressing \cite{PhysRevA.94.063825}.
Entangled photon pairs are proposed 
for ultrafast double-quantum-coherence spectroscopy
of excitons with entangled photons \cite{PhysRevA.82.013820} 
or quantum gates based on entanglement swapping protocols
\cite{PhysRevB.90.245419}.
The Schmidt decomposition allows
to analyze the material response function to obtain
information about otherwise inaccessible resonances of 
a complex system \cite{1510.06726}. This connects to the
context of quantum optical spectroscopy \cite{PhysRevA.73.013813,Hall:89} 
and nonlinearity sensing via photon-statistics excitation spectroscopy \cite{PhysRevA.84.053806,PhysRevB.79.035316}.
A very convenient method to simulate quantum excitation 
experiments is the quantum cascade setup developed at the 
same time by Gardiner \cite{PhysRevLett.70.2269}
and Carmichael \cite{PhysRevLett.70.2273}.
The quantum cascade approach allows a self-consistent mapping
of the quantum excitation onto a second-system, via the 
quantum Langevin \cite{PhysRevLett.70.2269} or quantum 
stochastic Schr\"odinger equations \cite{QN,carmichael2009open}.
This mechanism is a dissipatively mediated excitation
process as the output (measurement) of the source 
system is the input (excitation) of the target system.
This excitation strategy differs strongly from a bath
input (thermal equilibrium) or laser excitation, which 
adds coherence to the system.
In contrast, the quantum cascaded driving allows a 
photon-statistical fine tuning in between these regimes 
and renders a transient regime accessible, where 
thermal statistics and quantum coherences coexist
and intertwine via quantum emitters.
In this work, we theoretically discuss this intermixing 
and transition dynamics by employing a quantum cascaded
system.
In Sec.~\ref{sec:model}, we derive the basic quantum cascaded
coupling in the master equation formalism, equivalent to 
the method of Langevin operators \cite{PhysRevA.31.3761,PhysRevA.94.063825} 
and the quantum stochastic Schr\"odinger equation \cite{PhysRevLett.70.2273,PhysRevLett.116.093601}.
We apply this quantum cascaded coupling in Sec.~\ref{sec:cascade} to
a specific example: an incoherently pumped single quantum emitter
in a cavity as the source and as the target 
one or two identical quantum emitters coupled to a cavity.
We show that the intensity-intensity correlation $ g^{(2)}(0)$
of the target system follows the intensity-intensity correlation 
of the source in the regime of interest, however, classically 
degraded due to the mediating bath.
In Sec.~\ref{sec:hocorr}, we show that the response of the target system
follows not universally the output of the source system:
Higher-order intensity correlations $ g^{(n)}(0)$
exhibit a completely different picture.
Via these higher-order correlations, we finally discuss the 
qualitatively different behavior of the cascaded setup in comparison
to the typical excitation scenarios of coherent and incoherent
pumping in Sec.~\ref{sec:prop}.
%
%
%
In Sec~\ref{sec:conc}, we conclude and summarize the 
findings.

\section{Quantum cascade model}
\label{sec:model}

To investigate the dynamics of a quantum cascaded system, 
we derive a master equation in the Born-Markov limit 
\cite{PhysRevA.94.063825,PhysRevA.31.3761,PhysRevA.30.1386,*PhysRevLett.56.1917}.

We will consider systems as depicted in Fig. 
\ref{fig:scheme_example}
,
i.e. a source quantum system with Hamiltonian $ H_s $ coupled via a thermal 
bath, $ H_c $ to target quantum system $ H_t$.
The full Hamiltonian reads: $ H = H_0 + H_s + H_c + H_t$,
with $ H_0 $ including the free evolution dynamics of all 
quantities in the total system.
At this point, we do not define $ H_s $ and $ H_t $, but focus
on the coupling Hamiltonian $ H_c $, which is given in a rotating 
frame in correspondence to $ H_0 $ and reads in the rotating wave
approximation:
\begin{equation}
\frac{H_{\mathrm{c}}}{\hbar}= 
\int d\omega \ b(\omega) 
\Big[ K^s_\omega J_s^{\dagger}(t)   
+K^t_\omega J^{\dagger}(t,\tau)
\Big]  + \text{H.c.},
\label{eq:waveguide}
\end{equation}
where $ \tau $ describes the finite time delay between
the target and source system and $ J_s,J_t $ describe 
a single operator or a superposition in the source 
and target system, respectively.
The coupling of the source/target system to the connecting
reservoir is $ K^{s/t}_\omega$, which we set independent
of the frequency in the narrow bandwidth limit 
$ K^{s/t}_\omega \equiv K^{s/t}_0$.
To derive the quantum cascaded couling, we employ the canonical 
derivation of the master equation in the Born-Markov limit with 
$\chi_{\mathrm{tot}}(t)= \rho(t)\rho_B(0)$, assuming the coupling
reservoir in equilibrium and in a thermal state 
\cite{OpenQuantumSystems,statmeth1}:
\begin{align}
\frac{d\rho}{dt}|_c =
-\frac{1}{\hbar^2} \int_0^t ds 
\mathrm{Tr}_B 
\left\lbrace 
\left[ H_{\mathrm{c}}(t),\left[H_{\mathrm{c}}(s), \rho(t) 
\rho_B \right]\right] 
\right\rbrace
\label{eq:RF}.
\end{align}
We assume the thermal bath to be in the vacuum state and 
consider only contribution proportional to
$\ew{b(\omega)b^\dg(\omega)}$ and assume the 
commutator relations $ 
[b(\omega),b^\dg(\omega^\prime)]=\delta(\omega-\omega^\prime) $.
Given these conditions, the double commutator can be 
evaluated, and we yield the following master equation after
tracing out the bath degrees of freedom:
\begin{align} 
\nonumber
\frac{d\rho }{dt}|_c &=
-2\pi
\sum_{i=s,t}
(K^i_0)^2
\int_{0}^{t} ds \delta(s-t)
\\\nonumber
&\times\left[ 
J_i^\dg(t)J_i^\ndg(s)\rho(s)
-
J^\ndg_i(t)\rho(s)J^\dg_i(s)
-
J^\ndg_i(s)\rho(s)J^\dg_i(s)
+
\rho(s)J_i^\dg(s)J_i^\ndg(t)
\right] \\
\nonumber
\ & 
-
2\pi
K^s_0K^t_0
\int_{0}^{t} ds \delta(s-(t-\tau))\\\nonumber
&\times\left[ 
J_t^\dg(t)J_s^\ndg(s)\rho(s)
-
J^\ndg_t(t)\rho(s)J^\dg_s(s)
-
J^\ndg_s(s)\rho(s)J^\dg_t(t)
+
\rho(s)J_s^\dg(s)J_t^\ndg(t)
\right]
\\
\ & 
-
2\pi
K^s_0K^t_0
\int_{0}^{t} ds \delta(s-(t+\tau))\nonumber\\
&\times\left[ 
J_s^\dg(t)J_t^\ndg(s)\rho(s)
-
J^\ndg_s(t)\rho(s)J^\dg_t(s)
-
J^\ndg_t(s)\rho(s)J^\dg_s(s)
+
\rho(s)J_t^\dg(s)J_s^\ndg(t)
\right].
\end{align} 
We take into account that $\int^t_0 ds \delta(t-s) h(s)=h(t)/2$
and that $ s \le t$. By defining $ K^i_0=\sqrt{\gamma_i/(2\pi)}$,
where $\gamma_i$ are the decay rate of the subsystems that couple source and target.
Then, one coupling contribution between target and source vanishes.
The full master equation in the Born-Markov limit reads:
\begin{align}
\frac{d\rho }{dt} &=
\frac{1}{i\hbar} [H_s+H_t,\rho] \nonumber\\ \notag
&
+ 
\sum_{i=s,t}
\frac{\gamma_i}{2} \left(2 J^\ndg_i(t)\rho(t) J^\dg_i(t) 
- \lbrace J^\dg_i(t) J^\ndg_i(t),\rho(t)\rbrace \right) \\ \notag
&
-\sqrt{\gamma_s\gamma_t}
\left( 
J_t^\dg(t)J_s^\ndg(t_D)\rho(t_D)
-
J^\ndg_t(t)\rho(t_D)J^\dg_s(t_D) \right)
\\ 
&
-\sqrt{\gamma_s\gamma_t}
\left( 
\rho(t_D)J_s^\dg(t_D)J_t^\ndg(t)
-
J^\ndg_s(t_D)\rho(t_D)J^\dg_t(t)
\right),
\end{align}
with $ t_D=t-\tau$.
In our setup, the delay $ \tau $ is small and can be set to 
zero safely within our Markovian approximation.
Transforming back from the rotating frame, the full master
equation reads:
\begin{align} \notag
\frac{d\rho }{dt} &=
\frac{1}{i\hbar} [H_0+H_s+H_t,\rho] \\ \notag
&
+ 
\sum_{i=s,t}
\frac{\gamma_i}{2} \left(2 J^\ndg_i\rho J^\dg_i 
- \lbrace J^\dg_i J^\ndg_i,\rho\rbrace \right) \\
&
-\sqrt{\gamma_s\gamma_t}
\left( 
[J_t^\dg,J_s^\ndg\rho]
+
[\rho J^\dg_s,J^\ndg_t] \right).
\end{align}
Given this result, we can investigate different kinds 
of systems and study the particular features of a quantum
cascaded driving.
To characterize the cascaded driving, we choose 
first a specific system and then propose the photon-photon 
correlation functions as a measure for coherence in the 
system.
We will see that the cascaded system(exhibits different regimes of excitation depending on the source excitation, however, the source state is not straight forwardly mapped to the target system) is remarkably 
much closer to a coherent driving setup, although 
it is of purely dissipative nature.

\section{Example: Coupled cQED - Systems}
\label{sec:cascade}
\begin{figure}
\centering
\includegraphics[width=0.5\linewidth]{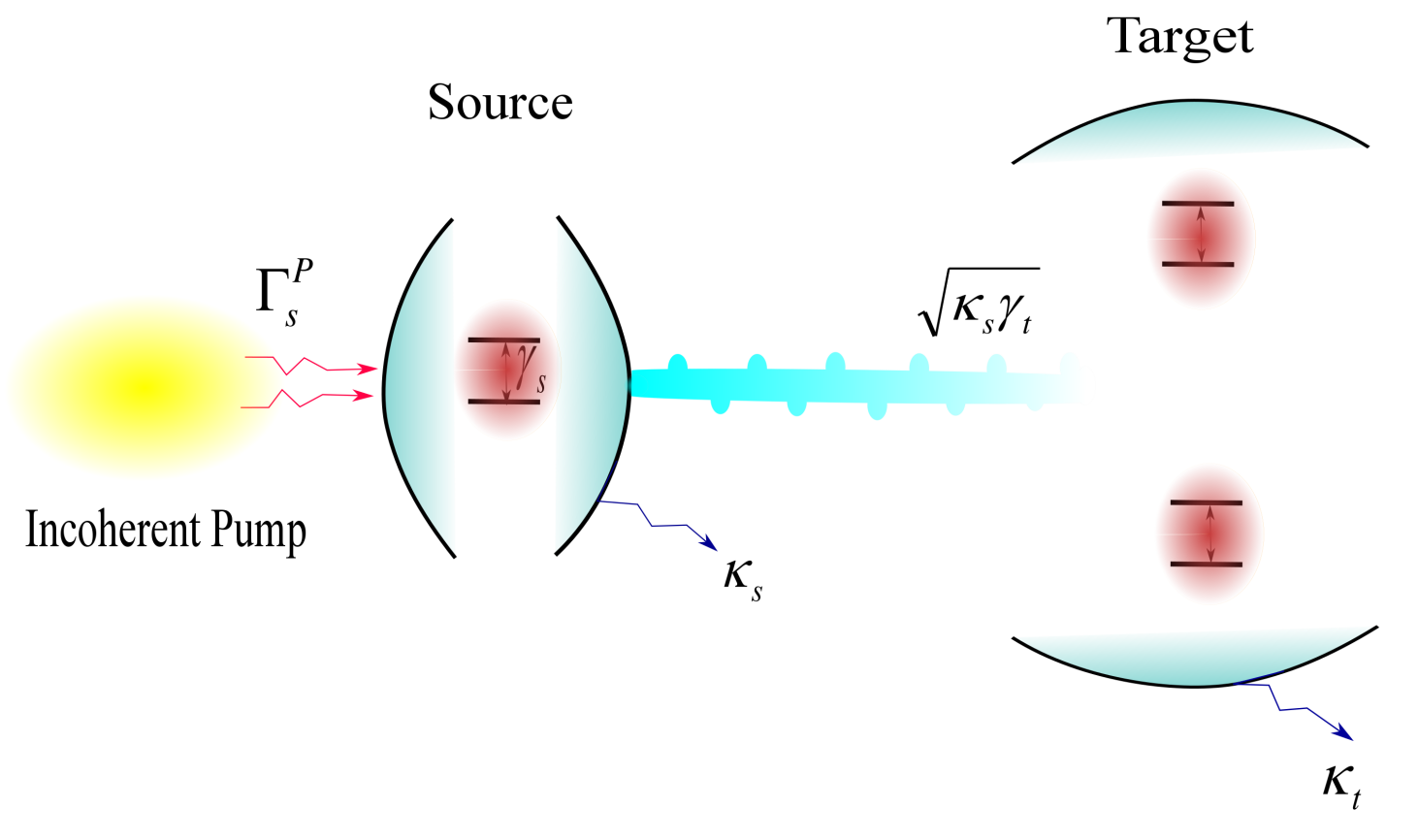}
\caption{Schematic depiction of the studied setup. The source cavity, which 
contains a TLS, is pumped incoherently with rate $\Gamma^P_s$. 
The emission of the  source cavity is fed into the one or two emitters contained in
the target cavity. }
\label{fig:scheme_example}
\end{figure}
As a platform to investigate quantum excitation in 
comparison to coherent and incoherent driving, we 
focus on a coupled cavity quantum electrodynamics (cQED) system, 
Fig.~\ref{fig:scheme_example}.
As a source system, we consider a single emitter 
coupled to a single cavity mode, which is the prototypical 
Jaynes-Cummings Hamiltonian:
\begin{equation}
H_{\mathrm{s}}= 
\hbar g_s 
\left( 
a_s^{\dagger} \sigma_s^- + \sigma_s^+ a_s 
\right),
\label{eq:source}
\end{equation}
where $ g_s=0.1 \text{ps}^{-1} $ denotes the coupling element between
the cavity field with creation(annihilation) operators $ a^{(\dg)} $
and the fermionic degrees of freedom, described via the 
spin Pauli matrices $ \sigma_s^{(+/-)}$.
The coupling operator from the source to the cavity is 
chosen to be $ J_s := a_s$.
To control the source system, we assume an incoherent pumping 
mechanism.
For far-off resonant driving, this pump mechanism can safely 
be described via \cite{del2010mollow,*PhysRevA.84.053804} 
\begin{align}
\mathcal{D}[\sqrt{\Gamma^P_s}\sigma^+_s]\rho 
:= 
\Gamma^P_s 
\left(
2\sigma_s^+\rho\sigma_s^- 
-
\lbrace \sigma_s^-\sigma_s^+,\rho\rbrace
\right),
\end{align}
assuming the transfer of excitation from the ground state to
the excited state of the fermionic system, and 
the definition 
$ \mathcal{D}[J]\rho :=2J\rho J^\dg - \lbrace J^\dg J,\rho\rbrace$.
In the following, we will fix all parameters 
(cf. Tab. \ref{tab:pars}) but 
$ \Gamma^P_s $, which 
is controllable via the intensity of the applied external pumping 
field, or even electrically steerable in semiconductor nanotechnology
platforms \cite{RIS_5,stevenson2006semiconductor,yuan2002electrically,schulze2014feedback}.

\begin{table}
\begin{tabular}{ l | c }
 parameter  & value ($\mathrm{ps}^{-1}$)  \\
 \hline
 $g_s$  & 0.1  \\
 $g_t$  & 0.1  \\
 $\gamma_s$  & 0.02  \\
 $\gamma_t$  & 0.5  \\
 $\kappa_s$  & 0.1  \\
 $\kappa_t$  & 0.005  
\end{tabular}
\caption{Parameters used for the cascaded setup throughout the manuscript.}
\label{tab:pars}
\end{table}

The source is an incoherently pumped single emitter coupled to 
a single cavity mode.
Depending on the pumping strength, the statistics of the output field
can be tuned over a wide regime, starting for weak 
pumping in the single-photon, or antibunching regime 
$g^{(2)}(0)=\ew{(a^\dg_s)^2(a^\ndg_s)^2}/\ew{a^\dg_s a^\ndg_s}^2<1$ via a 
synchronized laser transition $ g^{(2)}(0) \approx 1 $
to the thermal state for a pumping parameter $ \Gamma^P_s \gg g_s$.
To complete the picture, we assume a radiative decay for the 
source via 
\begin{align}
\mathcal{D}[\sqrt{\gamma_s}\sigma_s^-]\rho 
:= 
\Gamma^s_r 
\left(
2\sigma_s^-\rho\sigma_s^+ 
-
\lbrace \sigma_s^+\sigma_s^-,\rho\rbrace
\right).
\end{align}
The radiative decay amounts to $ \Gamma^s_r =0.02\text{ps}^{-1}$.
So, we assume not a perfect $ \beta=1 $ laser dynamics for the 
single emitter laser, as radiative decay is not fully absorbed 
by the cavity mode.
As a target system,
we choose also the Jaynes-Cummings Hamiltonian but with 
two emitters:
\begin{equation}
H_{\mathrm{t}}= 
\hbar 
\sum_{j=1,2}
g_{j,t}
\left( 
a_t^{\dagger} \sigma_{j,t}^- + \sigma_{j,t}^+ a_t 
\right),
\label{eq:target}
\end{equation}
where the emitter of the target system 
$\sigma_{j,t}^{-/+}$ couples to the single mode cavity 
with the strength of $ g_{j,t}=g_t=0.1\text{ps}^{-1}$
and the emitters are identical.
Here, the coupling operator from the target system
is chosen to be $ J_{i,t} := \sigma^-_{i,t}$ $(i=1,2)$, 
i.e. the coupling to the source is individual and not in 
superposition.
We assume an additional cavity loss for the target 
system via:
\begin{align}
\mathcal{D}[\sqrt{\kappa_t} a_t^\ndg]\rho 
:= 
\kappa_t 
\left(
2a^\ndg_t\rho a^\dg_t
-
\lbrace a^\dg_t a^\ndg_t ,\rho\rbrace
\right)
\end{align}
and setting the photon life time $ \kappa_t = 0.005\text{ps}^{-1}$.

The free evolution is governed by $ H_0 $ and given as:
\begin{align}
H_0 
=  
\hbar\omega_0 \sum_{i=s,t} a_i^{\dg} a^\ndg_i 
+
\hbar\omega_e
(
\sigma_s^+\sigma_s^- 
+
\sum_{i=1,2}
\sigma_{t,i}^+\sigma_{t,i}^-
), 
\end{align}
We assume a resonant dynamics between cavity and the 
emitter $ \omega_e=\omega_0 $ and also in between 
the source and target.
Therefore, the full master equation reads:
\begin{align} \notag
\label{eq:meq}
\frac{d\rho }{dt} &=
\frac{1}{i\hbar} [H_0+H_s+H_t,\rho] \\ \notag
&
+ 
\mathcal{D}[\sqrt{\Gamma^P_s}\sigma_s^+]\rho
+
\mathcal{D}[\sqrt{\gamma_s}\sigma_s^-]\rho 
+
\mathcal{D}[\sqrt{\kappa_s}a^\ndg_s]\rho
\\ \notag
&
+
\mathcal{D}[\sqrt{\kappa_t}a^\ndg_t]\rho
+
\sum_{i=1,2}
\mathcal{D}[\sqrt{\gamma_{t}}\sigma_{t,i}^-]\rho 
\\
&
-\sqrt{\kappa_s\gamma_t}
\sum_{i=1,2}
\left( 
[\sigma^+_{t,i},a^\ndg_s\rho]
+
[\rho a^\dg_s,\sigma_{t,i}^\ndg] \right).
\end{align}
This master equation is numerically evaluated with 
a fourth-order Runge-Kutta algorithm for different 
values of $ \Gamma^P_s $.
We keep throughout the 
discussion all other values fixed and cast the 
master equation Eq.~\eqref{eq:meq} into the 
basis $ 
\bra{e_s,p_s,e_t,p_t}\rho\ket{e_s^\prime,p_s^\prime,e_t^\prime,p_t^\prime}$
with $ e_i $ emitter states and $ p_i $ photon manifold of source and
target $ i=s,t $.
We compute the observables for different photon manifold
cut-offs $ p_i<N_i $ until convergence is reached, 
i.e. the corresponding and discussed observable do not 
change by increasing the cut-off further.
We restrict our discussion in the following
to observables of photon manifolds $ p_i\le 10 $.
%

\begin{figure}
\centering
\includegraphics[width=0.45\linewidth]{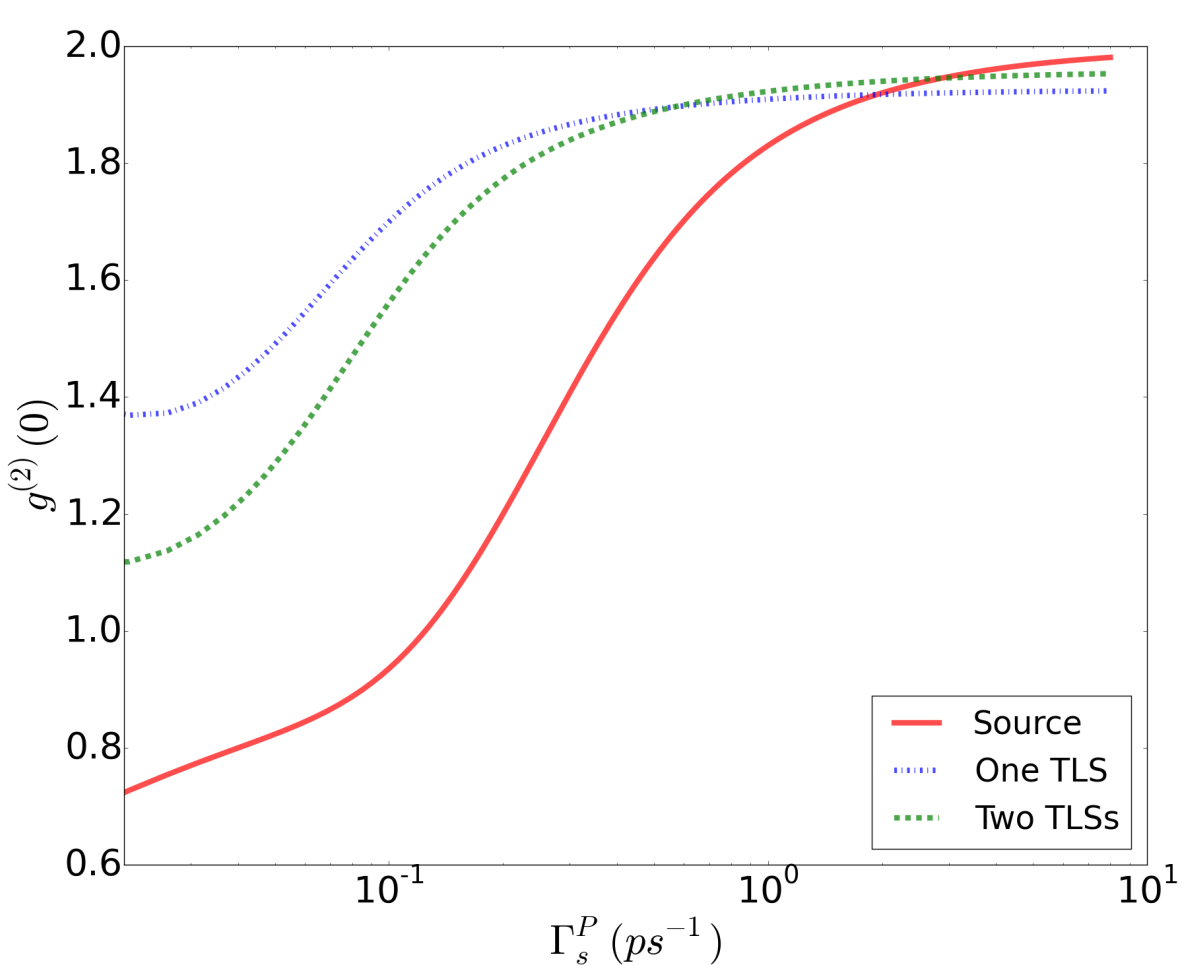}
\caption{Second order correlation functions $g^{(2)}(0)$ of source (red, solid) and one (blue, dashed dotted) and two (green, dashed) tLSs in the target cavity. For low pump rates 
the target, in contrast to the source, shows rather a bunching behavior. 
When increasing the pump strength,for both source and target, a transition to the thermal regime occurs.  }
\label{fig:g2}
\end{figure}

%
We discuss the response of the target system with respect to
the photon-statistics of the output field.
The output is included via the cavity loss of the 
target, and can be measured in Hanbury Brown and 
Twiss setups via
the second-order correlation function, defined in the 
steady state limit as \cite{loudon2000quantum}:
\begin{equation}
g^{(2)}_{\mathrm{stat}}(\tau)=\lim\limits_{t\rightarrow\infty}
\frac{
\left\langle 
a^{\dg}_i(t)a^{\dg}_i(t+\tau) a^{\ndg}_i(t+\tau) a^{\ndg}_i(t) 
\right\rangle}
{\left\langle a^{\dg}_i(t) a^{\ndg}_i(t) \right\rangle^2},
\label{eq:g2}
\end{equation}
where for the source $ i=s $ and for the target cavity $ i=t $. 
We consider here only the coincidence rates for zero delay $ \tau=0$,
as in this limit quantum effects in the correlation are prominent.
In Fig.~\ref{fig:g2}, we numerically evaluate the $ g^{(2)}(0) $
for the source (red, solid)  
and target with one (blue, dashed dotted) and two (green, dashed) TLSs 
for different incoherent pumping strengths $ \Gamma^P_s $:

The source can be driven into the antibunching regime $ g^{(2)}<1 $
for driving strengths of $ \Gamma^P_s<g_s$, where single photons are emitted.
The cavity coupling is not strong enough to produce more than
one cavity photon, before the cavity loss and dissipation forces
the photon to leave the resonator.
The source dynamics stays antibunched for a wide range of 
parameters and turns coherent for a pumping strengths $ \Gamma^P_s>g_s$
and for even larger pumping, the pumping induced dephasing
adiabatically eliminates the emitter dynamics and the output field 
equilibrates into a thermal state \cite{PhysRevA.84.053804,PhysRevA.50.4318,Ritter:10}. 
Focusing now on the target dynamics, 
we observe that the dissipative coupling via the reservoir
leads to a more classical response for one (blue, dashed dotted) as well as two TLSs (green, dashed line).
In comparison to the source statistics, even in the case of two quantum emitters, i.e.,
with a stronger quantum nonlinearity, the photon
statistics in target cavity is less non-classical.
We can explain this due to the dissipative transfer mechanism
between the cavities,
leading to thermal mixture and loss of coherence from the 
source to the target.
We observe furthermore that in the regime $\Gamma^P_s>g_s$,
the response follows the 
source dynamics, and we conclude that a quantum cascaded 
coupling does not qualitatively change the second-order 
photon correlation function of the target.
However, we will see that this is not the case for higher-order
photon correlation functions, which we discuss in the next section.
%

\section{Beyond the second order photon correlation}
\label{sec:hocorr}
\begin{figure}[h]
\centering
\includegraphics[width=0.5\linewidth]{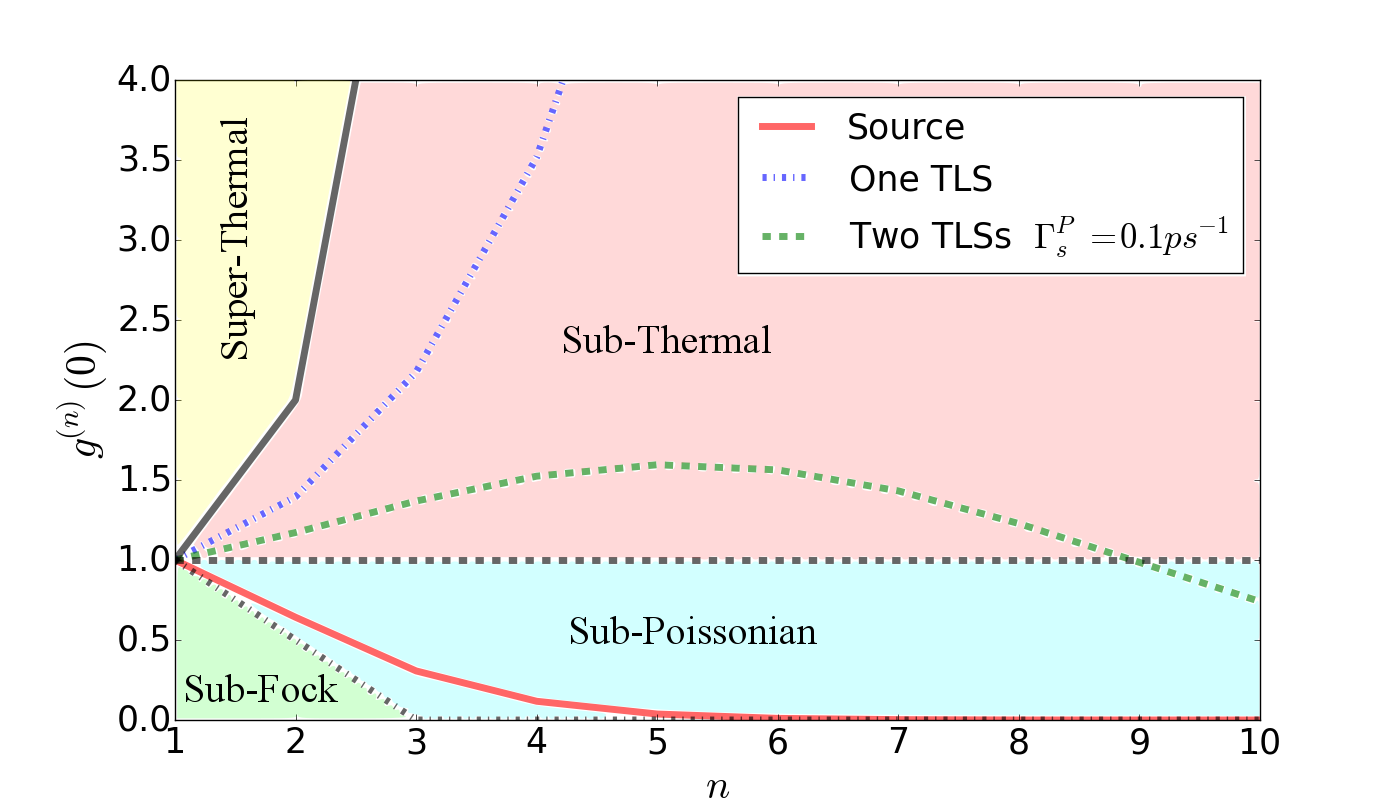}
\caption{
Correlation functions in the steady state, when the pump strength is equal to the cavity coupling ($\Gamma^P_s=g$) for source (red, solid) and target with two TLSs (green, dashed) and a single TLS (blue, dashed dotted). The solid, dashed and dash dotted gray lines present thermal, coherent and pure quantum light. The source is antibunched and in the subpoissonian regime for all orders in the correlation function. The target with a single TLS exhibits thermal light. However, the target cavity when containing two TLSs shows a transitional behavior, where it starts out in the sub-thermal regime but goes to the sub-Poissonian regime for higher orders.
}
\label{fig:gn}
\end{figure}
%
%
%

\begin{figure*}
\centering
\includegraphics[width=\linewidth]{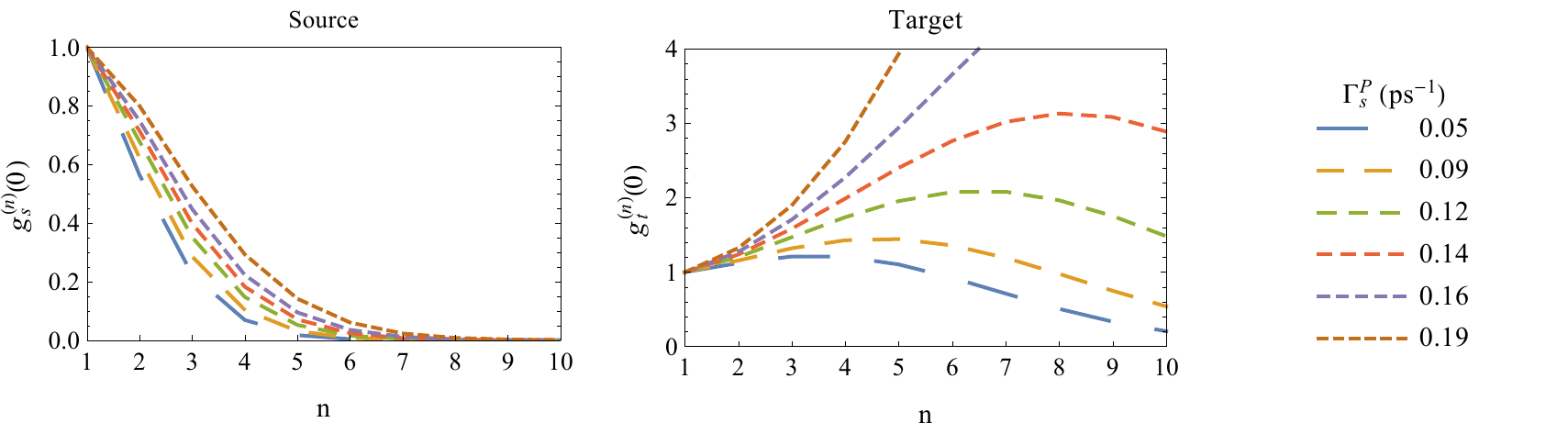}
\caption{
Higher-order correlation functions of the source $ g^{(n)}_s(0) $
and target system $ g^{(n)}_t(0) $ for different incoherent pumping
strength of the source system $ \Gamma^P_s $.
Remarkably, the target system exhibits a different behavior than 
the source system. 
}
\label{fig:gn_diff_pumping}
\end{figure*}

%
Experimentally, higher-order photon-correlations
have become accessible \cite{Assmann297}.
They allow to characterize the quantum light field
in photon detection experiments more precisely.
For example, a $g^{(2)}(0)\approx 1$ is often taken to be a
sign for a coherent 
light field (in the Glauber state), or a Fock state with 
a large photon number: $ g^{(2)}_\text{Fock}(0)=1-1/N
\rightarrow 1$ for $N=\ew{a^\dg a} \gg 1$.
However, only considering higher-order correlations
allows for a definite characterization of the light field.
These are defined for $\tau=0$ and in the steady-state as 
\begin{equation}
g^{(n)}_{\mathrm{stat}}(0)=
\frac{
\left\langle a^{\dg n}_i a^{\ndg n}_i \right\rangle}
{\left\langle a^{\dg}_i a^{\ndg}_i \right\rangle^n},
\label{eq:gn}
\end{equation}
where $ i=s $ for the source and $ i=t $ for the target cavity. 
Measuring such higher-order correlations allows to  
discriminate output fields even in case, when 
the $g^{(2)}(0)$ function value is equal. 
For example, the Fock state 
higher-order correlation functions read 
$ g^{(n)}_\text{Fock}(0)=N!/[N^n (N-n)!]$
for $n<N=\ew{a^\dg a}$ and therefore 
$ g^{(n)}_\text{Fock}(0)>g^{(n+1)}_\text{Fock}(0)$
in contrast to a coherent distribution, which holds
$ g^{(n)}_\text{coh}(0)=g^{(n+1)}_\text{coh}(0)=1$.
For a thermal light field with $ \bar n $ 
mean photon number, the unnormalized higher-order 
correlation functions read
$
\left\langle a^{\dg n} a^{n} \right\rangle
=n! (\bar n)^n
$ 
and are calculated from $ p_n= (\bar n)^n/(1+\bar n)^{n+1} $.
For the correlation function holds then 
$ g^{(n)}_\text{therm}(0)<g^{(n+1)}_\text{therm}(0)=(n+1)!$.
We take these three limiting cases, to visualize our 
quantum cascade driving setup.
In Fig.~\ref{fig:gn}, we plot the higher-order correlation
functions for the source cavity (red, solid) and the target
cavity with one (blue, dashed dotted) and two (green, dashed) TLSs.
To illustrate regimes, we shaded the areas that distinguish
between super-thermal and sub-thermal fields, and super- and sub-Fock
states.
The Fock state limits are taken, so that the number of Fock 
photons equals the order of the correlation function $N=n$.
The correlations of the source $ g^{(n)}_s(0) $ and target system 
$ g^{(n)}_t(0)$ are shown for $ \Gamma^P_s=0.1\text{ps}^{-1}=g$. 
The output field of the source shows a monotonic behavior,
i.e. $ g^{(n)}_s>g^{(m)}$ for all $ n<m<10 $.
That is, comparing to the shaded area, very characteristic
for a non-classical output field.
For one TLS in the target cavity, we also see a monotonic behavior,
but this time with $ g^{(n)}_s<g^{(m)}$ for all $ n<m<10 $.
In contrast, the output field of the target with two TLSs does not 
exhibit this kind of monotonous behavior, e.g. 
$ g^{(2)}_t(0)<g^{(3)}_t$ but $ g^{(2)}_t(0)>g^{(6)}_t$.
Given this difference, it is clear, that the target
dynamics is not a simple image of the source, the differences
cannot only be traced back to dissipative degradation
from the source-target transfer. In the following, we will focus on
the case of two TLSs, as we find interesting photon probability 
distributions for this case, and also, since the target system
with one emitter does not show non-monotonic behavior in the 
parameter regime, in which we are interested in.
The quantum cascade coupling introduces an own, remarkable
behavior and prevents a straightforward imprinting
of the source statistics on the target quantum statistics, i.e. the $ g^{(n)}_t(0) $
distribution.

In Fig.~\ref{fig:gn_diff_pumping}, we investigate the
higher-order correlation functions of the source $ g^{(n)}_s(0) $
and target system $ g^{(n)}_t(0) $ for different incoherent pumping
strengths of the source system $ \Gamma^P_s $.
Interestingly, the response of the target differs strongly
from the source quantum statistics.
The source system shows a monotonic behavior for all pumping strengths:
$ g^{(n)}_s(0)> g^{(m)}_s(0) $ for all $n<m\le10$.
Furthermore, the quantum statistics approaches lower values and 
reaches small values for high orders.
This behavior is expected, since the incoherent driving and 
the cooperativity \cite{nla.cat-vn4402306} $ C_s=g_s^2/(\Gamma_R\kappa_s)$ 
limits the 
achievable photon manifold, i.e. there is always a cut-off $ n_c $ with
$ p_{n_c}=0 $ and therefore the importance of higher-order 
correlations decreases: $ g^{(n)}_s(0)\rightarrow0 $ for 
$ (n-n_c) \rightarrow 0 $.
In contrast, the target system reaches first a maximum for a certain
$ m $ with $ g^{(m)}_t(0) \ge g^{(n)}_t(0) $ for all $ n $.
This maximum shifts, as expected, for higher pumping strength 
towards larger $ m $, since the maximum number of photons also 
shifts to larger values. 
After the maximum, the $g_t^{(n)}(0)$ distribution follows the trend of the 
source system towards lower values.
This behavior is stable for a wide range of pumping strengths.
Due to the presence of a cut-off in the source photon
manifold $ n_c $, the target quantum distributions will
also, eventually, tend to zero.
However, the target system follows only for large $ n $, 
the source  quantum statistics, always after passing a maximum.
This maximum, however, can shift to very large values, and 
in particular from a certain pumping strength on: 
$ \Gamma^P_s=0.2\text{ps}^{-1}$ (blue, dashed line).

\begin{figure}
\centering
\includegraphics[width=0.45\linewidth]{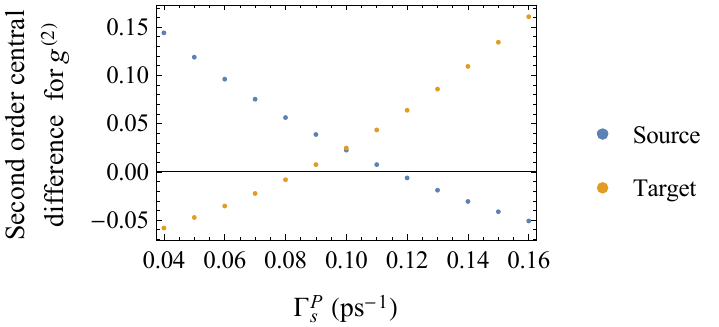}
\caption
{The transition observed in the system illustrated by the second order finite difference at the $g^{(2)}$-function. While the source correlations cross from an upwards to a downwards turning point, the target correlations exhibit the opposing behavior. The curves cross at the coupling strength $g=0.1 \mathrm{ps}^{-1}$ common to source and target.}
\label{fig:transition}
\end{figure}
Furthermore, Fig. \ref{fig:gn_diff_pumping} shows a qualitative transition of the target system in the correlation functions.
For low incoherent pump strengths, the curve is turning downwards. Then there is a transition towards the regime, where the curve turns upwards. We quantify this by the second order central difference defined as
\begin{equation}
g^{(n)\prime\prime}=\frac{g^{(n+1)} - 2 g^{(n)} + g^{(n-1)} }{(n+1-n)(n-(n-1))}.
\end{equation}
During the transition from coherent to thermal behavior the $n$th order correlation function will flip successively up.
Here, we characterize this transition by the second order difference at the $g^{(2)}$-function, which will first show the flip, so that the curve points here upwards.
This is shown in Fig. \ref{fig:transition}, where we observe, that the target system goes from a downwards to an upwards
turning point. At the same time the source system shows a transition from an upwards to a downwards turning point.
The curves cross at the coupling strength $g=0.1 \mathrm{ps}^{-1}$. Thus, even though it is not straightforwardly
obvious how the source influences the target, we can illustrate the transition in the target system by a corresponding transition in the source system

To explain the origin of our results, in the next section, we compare the quantum statistics
of the target system with a coherent and incoherent 
drive.
We will see, that this maximum in the photon-correlation
is not readily produced with either coherent or incoherent driving.
Thus, the cascaded setup allows to create photon statistics not
achievable with a reduced formulation.

\section{Properties of cascaded driving}
\label{sec:prop}
%
\begin{figure}
\centering
\includegraphics[width=0.25\linewidth]{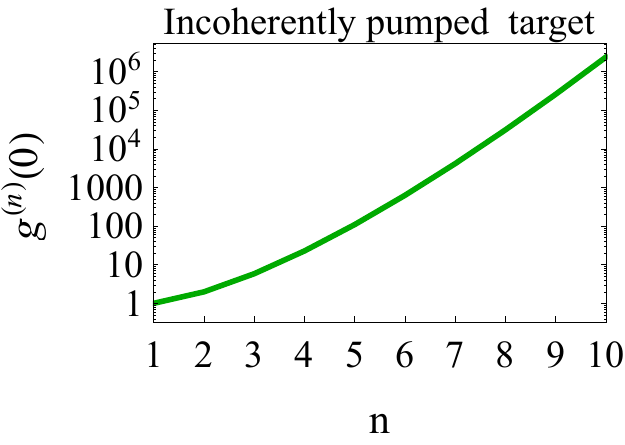}
\includegraphics[width=0.25\linewidth]{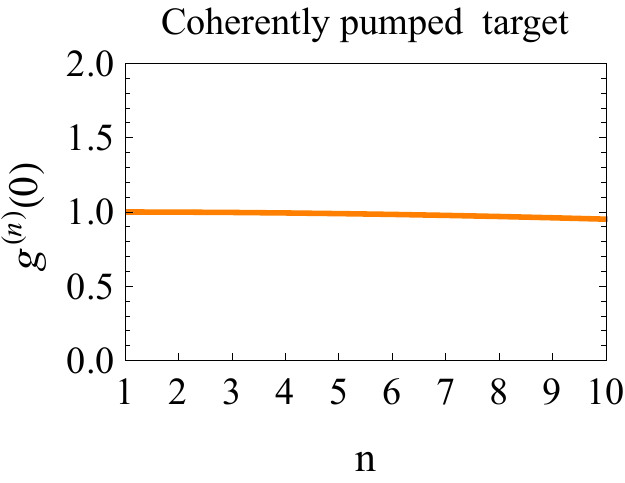}
\caption
{
Higher-order correlation functions of the target system 
with no quantum source pumping $ \kappa_t=\Gamma^P_s=0$.
Instead the target is directly pumped incoherently (left) and coherently (right) with 
$ \Gamma^P_t=\sqrt{\gamma_t\kappa_s}$. Note the logarithmic scale for incoherent pumping, and the 
monotonous increase in contrast to the coherent driving induced
maximum in the $ g^{(n)}(0) $ distribution. The incoherent driving exhibits thermal statistics,
while the coherent driving is close to coherent statistics for a wide range of pump parameters. }
\label{fig:gn_coh_incoh}
\end{figure}
To characterize the quantum cascade, we compare the 
resulting higher-order correlation with a system that
is coherently or incoherently driven.
To model this situation, we switch the coupling
between the source and target system off by setting $ \kappa_s=0$.
The driving of the target system is now included for 
the coherent driving by displacing the target's
photon operator according to 
$ a^\dg_t \rightarrow a^\dg_t + \Gamma^P_t/g_t  $
and for the incoherent driving case, we switch 
the operators of the incoherent pumping from 
$ \mathcal{D}[\sqrt{\Gamma^P_s}\sigma^+_s]\rho
\rightarrow \mathcal{D}[\sqrt{\Gamma^P_t}\sigma^+_t]\rho$.

\begin{figure}
	\centering
	\includegraphics[width=0.35\linewidth]{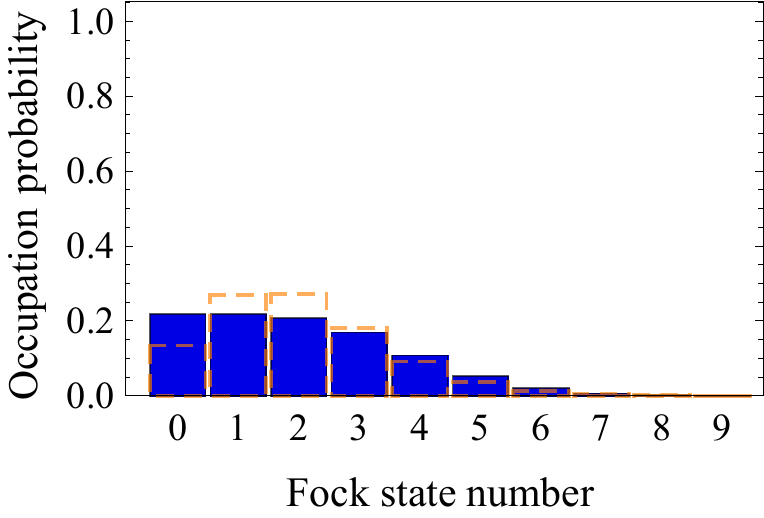}
	\caption{Occupation probability of the Fock states for
		$\Gamma^P_s=0.1\mathrm{ps}^{-1}$ corresponding to the
		photon statistics shown in Fig. \ref{fig:gn} (solid, blue). Due to
		the cascaded coupling the photon number distribution
		is exceptionally flat. This illustrates the photon
		statistics that deviate from the prototypical cases.
		For reference (dashed, orange), the coherent distribution
		is shown.}
	\label{fig:photocc}
\end{figure}

In Fig.~\ref{fig:gn_coh_incoh}, we compare the 
higher-order correlation functions for the case
of coherent pumping (left panel) and incoherent 
pumping (right panel) of the target system.
All parameter values are kept to allow comparison
with the quantum cascaded case.
Comparing the behavior of the correlation functions
in the cascaded setup (cf. Fig. \ref{fig:gn_diff_pumping}),
with the incoherently and the coherently pump cases,
we see a qualitatively different behavior.
While, the cascaded setup exhibits a maximum in the
correlation functions, the incoherently driven one
exhibits thermal behavior, increasing monotonically
and the coherently driven system exhibits close to coherent
statistics.
The form of the photon statistics for the cascaded system
is distinctly different than for the other excitation
scenarios.
This is consistent with the findings in Ref. \cite{PhysRevA.94.063825},
where it is shown that, in principle, the target of a stationary cascaded
system may access parts of the Hilbert space, that 
would not be accessible by other means.
Here, we illustrate this finding by showing a 
physical system realizing this possibility.

If we inspect the coupling terms, we can give 
some physical intuition for the observed result.
While the cascaded coupling is derived using an intermediate
bath and thus constitutes a dissipative coupling,
the coupling preserves some properties of the source statistics
in certain regimes.
This becomes clear from the master equation Eq.~\eqref{eq:meq}.
If one exchanges $ \sqrt{\gamma_t\kappa_s} 
\rightarrow -\sqrt{\gamma_t\kappa_s}$, the system 
dynamics and results remain unchanged, as it is 
the same with $ H_{t/s} \rightarrow -H_{t/s} $.
This explains the part of the dynamics that preserve
the source photon statistics for low pump strengths.
This behavior is not expected from a dissipative coupling
as the standard Lindblad form is independent of a change in the sign.
For weak incoherent pumping, quantum coherences can 
be built up and those quantum processes are mediated
via $ a^\dg_s $ to the  coherences of the target system
$\sigma^+_t$.
In this limit, for high pumping strengths, the system becomes thermal.
However, the intermediate coupling regime shows the
transition, allowing for peculiar distributions by only partially imprinting the source photon statistics on the target in the high-order correlation functions.
The Fock distribution corresponding to the statistics in
Fig. \ref{fig:gn} is shown in Fig. \ref{fig:photocc}.
Here, we observe a very flat distribution exhibiting a similar
probability for the first few photon number states(solid,blue).
This deviates from the coherent distribution(dashed,orange), which exhibits a
maximum and the thermal distribution, which decreases monotonically.
With this, we can explain the accessibility of new photon statistics
by the mixture of Hamiltonian and decoherent coupling processes, mediated by the cascaded setup.

\section{Conclusion}
\label{sec:conc}
%
We investigated a quantum cascaded system, in which
an incoherently pumped source system drives a target 
system with its quantum output field.
As observables, we focused on higher-order photon-
correlations $ g^{(n)}(0)$.
We find that the response of the target system differs 
strongly for different values of the incoherent pump 
parameter. 
For low values in comparison to the coupling constant of
the target system $ \Gamma^P_s<g_t$, the quantum 
statistics of the source system are imprinted on the target system.
For larger values the target system's 
output field resembles an incoherently driven quantum system.
However, in an intermediate regime, a mixture of coherent
and incoherent processes due to the coupling mechanism occurs
leading to quantum statistics differing from the prototypical
coherent and thermal shapes and giving rise to the possibility
of producing flat photon distributions.

\begin{acknowledgments}
S.C.A. thanks DAAD foundation for the visiting research grants. 
N.L.N., A.K., and A.C. are grateful towards the Deutsche Forschungsgemeinschaft for support
through SFB 910 ‘‘Control of self-organizing nonlinear systems’’ (project B1).
N.L.N also acknowledges support through the School of Nanophotonics (Deutsche Forschungsgemeinschaft SFB 787).

\end{acknowledgments}

\bibliographystyle{apsrev4-1}
%

\end{document}